\documentclass[twocolumn,showpacs,preprintnumbers,amsmath,amssymb]{revtex4}

\usepackage[dvips]{graphicx}
\usepackage{dcolumn}
\usepackage{bm}
\begin{document}

\preprint{Accepted for publication in Applied Surface Science}

\title{Scanning tunneling microscopy and spectroscopy studies of graphite 
edges}

\author{Y. Niimi}
\author{T. Matsui}
\author{H. Kambara}
\author{K. Tagami}
\author{M. Tsukada}
\author{Hiroshi Fukuyama}
 \email{hiroshi@phys.s.u-tokyo.ac.jp}
\affiliation{%
Department of Physics, University of Tokyo, 7-3-1 Hongo Bunkyo-ku, Tokyo 113-0033, Japan
}%

\date{April 5, 2004}

\begin{abstract}
We studied experimentally and theoretically the electronic local density of 
states (LDOS)
near single step edges at the surface of exfoliated graphite.
In scanning tunneling microscopy measurements, we observed the $(\sqrt{3} 
\times \sqrt{3}) R 30^{\circ}$
and honeycomb superstructures extending over 3$-$4 nm both from the 
zigzag and armchair edges.
Calculations based on 
a density-functional derived non-orthogonal tight-binding model show
that these superstructures can coexist if the two types of edges admix each 
other
in real graphite step edges.
Scanning tunneling spectroscopy measurements near the zigzag edge reveal a 
clear peak in the LDOS
at an energy below the Fermi energy by 20 meV.
No such a peak was observed near the armchair edge.
We concluded that this peak corresponds to the ``edge state" theoretically 
predicted for graphene ribbons, since a similar prominent LDOS peak due to 
the edge state is obtained by the first principles calculations.
\end{abstract}

\maketitle

\section{Introduction}
Graphite is one of the most extensively studied materials both 
experimentally and theoretically.
Recently, electronic properties of graphite nanostructures, such as graphite 
edges, have been attracting much attention in view not only of basic research 
but also of applications.
For example, porous carbons which consist an assembly of minute 
graphite fragments
and naturally of a large amount of graphite edges have extremely high 
specific surface area~\cite{porous,fragments}.

\begin{figure}
\begin{center}
\includegraphics[width=7cm]{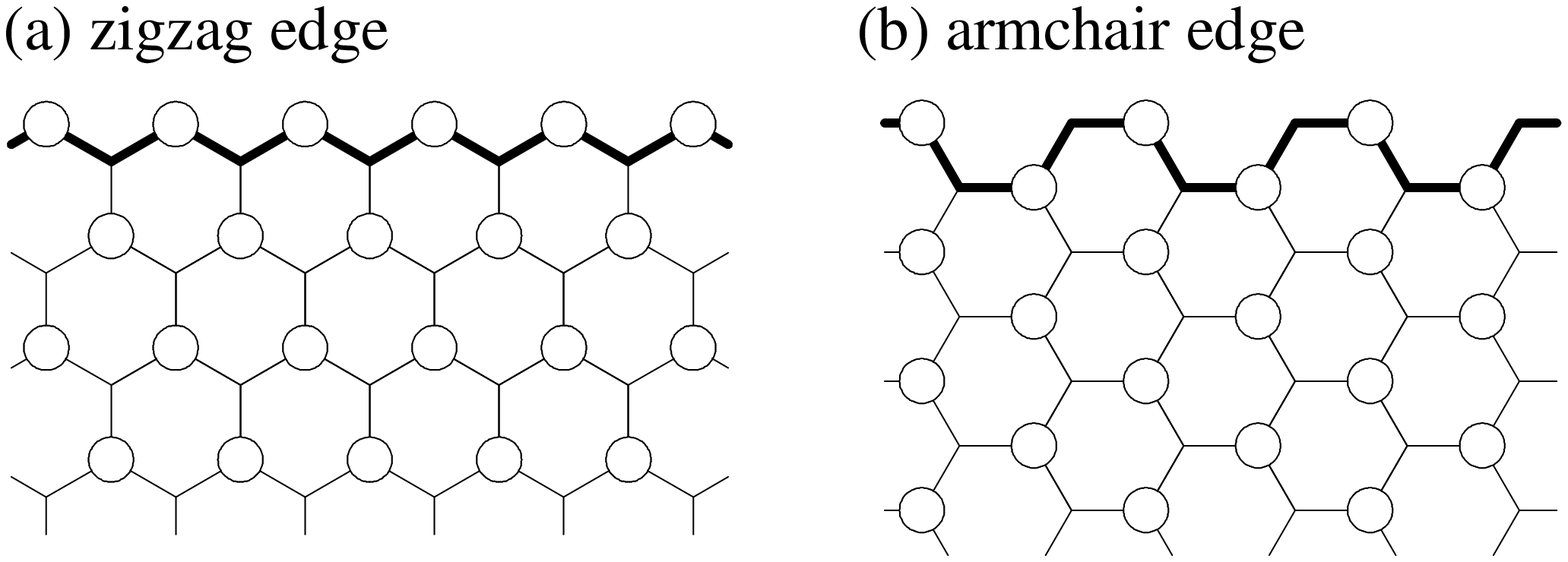}
\caption{Two graphite edges represented by the bold 
lines: zigzag edge (a)
and armchair edge(b).
The open circles are B-site carbon atoms.}
\label{edge_structure_fig}
\end{center}
\end{figure}

Topologically, there are two types of edges in single-layer graphite sheet 
(graphene),
i.e., zigzag and armchair edges (see Fig.~\ref{edge_structure_fig}).
Fujita \textit{et al}.~\cite{fujita1} were the first to predict the 
existence of
the specific electronic states localized only at the zigzag edge
from the tight binding band calculations for graphene ribbons.
On the other hand, it does not appear at the armchair edge.
The flat band nature of this ``edge state" results in a peak in
the local density of states (LDOS) at the Fermi energy ($E_F$).
When the ribbon width is large enough, the contribution of the edge state
to the total density of states should be negligibly small.
However, this could be measurably large if we survey the LDOS, for example,
with the scanning tunneling spectroscopy (STS) technique in the vicinity of 
zigzag edge.
It is also predicted that a similar edge state appears at zigzag edges
of multi-layer $\alpha \beta$ stacking ribbons~\cite{miyamoto}.
This suggests that the edge state should be potentially observed at step 
edges on bulk graphite surfaces.

In the previous STS measurements \cite{klusek1,klusek2}, a broad maximum 
near $E_F$
in the tunnel spectrum was reported to appear near circular edges of 
graphite nanopits.
This broad maximum might be attributable to the edge state.
However, at a circular edge around the nanopit, both the zigzag and 
armchair edges
inevitably coexist with comparable probabilities, which makes it difficult 
to extract
an electronic property for either of the edges.
In another scanning tunneling microscopy (STM) measurement \cite{armchair_stm},
a $(\sqrt{3} \times \sqrt{3})R 30^{\circ}$ superstructure was observed near 
an armchair edge
at a surface of highly oriented pyrolytic graphite (HOPG).
So far, it is not clear whether similar superstructures appear near the 
zigzag edge or not \cite{zigzag_stm}.

In this report, we present results of STM and STS measurements
for single step edges of both the zigzag and armchair types on a surface of 
exfoliated graphite, 
which contains much higher edge densities than non-exfoliated HOPG.
With STS, we observed a peak in a tunnel spectrum at an energy just below $E_F$
on a terrace with the zigzag edge but not on that with the armchair one.
This is the first clear spectroscopic observation of the theoretically 
predicted
graphite edge state \cite{fujita1,miyamoto}.
We also found two types of superstructures, i.e., the $(\sqrt{3} \times 
\sqrt{3})R 30^{\circ}$
and honeycomb ones, coexisting on the terrace with both zigzag and 
armchair edges.
Our calculations based on  
the density-functional derived non-orthogonal tight-binding model 
show that this coexistence is due to admixing of 
the two types of edges on the exfoliated graphite surface.

\section{Experimental}
All the measurements shown here were made on ZYX exfoliated graphite 
(hereafter ZYX)~\cite{niimi}.
ZYX was made from HOPG by graphite intercalation with HNO$_3$ and 
by subsequent
evacuation of the intercalant at high temperature. 
And then, it was evaculated at 1500 $^{\circ}$C for 3 h to remove 
the remnant intercalants.
The single crystallite size (100$-$200 
nm~\cite{birgeneau,niimi1}) is smaller
than that in HOPG by an order of magnitude, 
which makes it much easier to find step edges on the surface 
with STM.
All the graphite edges studied are monoatomic in height
with an almost linear shape in the length scale of 100 nm or longer.
We believe that active $\sigma$-orbital bonds at the edges are terminated 
by hydrogen
or else since we did not try to remove them in ultra high vacuum (UHV) 
at elevated temperatures.
Other details on the characterization of ZYX have been published 
elsewhere~\cite{niimi}.

The STM images were obtained at room temperature in air with mechanically 
sharpened
Pt$_{0.8}$Ir$_{0.2}$ wire tips.
The data were acquired in the constant current mode with a tunnel current 
($I$) of 1.0 nA
and a bias voltage ($V$) of $+0.1$ V applied to the sample with respective 
to the tip.
The STS data were obtained at $T = 77$ K in UHV of $P \leq 2 \times 10^{-7}$ Pa
with our newly constructed STM~\cite{ULT-STM}.
A tunnel spectrum was obtained by averaging a set of $dI/dV$ vs. $V$ 
curves measured at
100 to 120 grid points over $5 \times 5$ to $10 \times 10$ nm$^2$ area
by the lock-in technique ($f = 71.73$ Hz, $V_{{\rm mod}} = 6$ mV).

\section{Theoretical models}
In the calculations we assumed two layers of graphene 
for both types of edge structures.
The bottom layer is composed of a graphene  
extended infinitely over the two dimensional plane, 
and is assumed to have no defects.
The top layer with edge structures 
is prepared by removing a few tens to hundreds atoms 
from an infinite graphene.
The lateral dimensions of the periodic supercells in the top layer
are 12.8 $\times$ 14.8 nm$^2$ and 7.4 $\times$ 25.6 nm$^2$
for the zigzag and armchair edges, respectively.

The electronic states of these graphite layers are calculated
by the density-functional derived non-orthogonal tight-binding model
\cite{Frauenheim98}.
We assumed that carbon atoms at the edges are completely hydrogen terminated, 
and took into account only the $\pi$-orbital 
at each carbon atomic site.
These are relevant to the electronic states around $E_{F}$.
LDOS at every atomic site
is obtained by diagonalizing
the Hamiltonian and overlap matrices generated from these $\pi$-orbitals 
at the $\Gamma$ point.

\section{Results and discussions}

\begin{figure}
\begin{center}
\includegraphics[width=4.8cm]{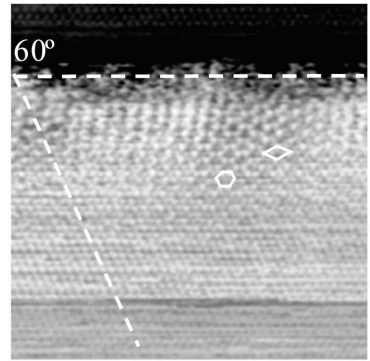}
\caption{An STM image near a zigzag step edge (the dashed line) 
on the surface of ZYX exfoliated graphite (8 nm $\times$ 8 
nm, $T=300$ K, in air).
The long dashed line shows the atomic row of B-site atoms.
The diamond and hexagon represent the $(\sqrt{3} \times \sqrt{3}) R 
30^{\circ}$ superstructure
and the honeycomb one, respectively.}
\label{zigzag_superstructure_fig}
\end{center}
\end{figure}

\begin{figure}
\begin{center}
\includegraphics[width=4.8cm]{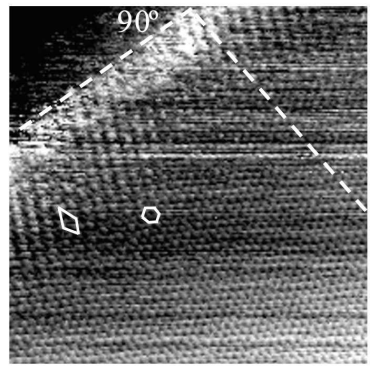}
\caption{An STM image near an armchair edge 
(the dashed line)
on ZYX (8 nm $\times$ 8 nm, $T=300$ K, in air).
The long dashed line shows the atomic row of B-site atoms.
The diamond and hexagon represent the same meanings as in 
Fig.~\ref{zigzag_superstructure_fig}.}
\label{armchair_superstructure_fig}
\end{center}
\end{figure}

An STM image taken near a zigzag step edge on the ZYX surface is shown in 
Fig.~\ref{zigzag_superstructure_fig}.
Although an atomic resolution is not obtained right on the edge,
we can identify it as the zigzag type by noting that the atomic row of 
B-site carbon atoms
(the long dashed line) on the upper terrace (the lower part of the 
image) is oriented at 60$^{\circ}$
to the edge direction (the dashed line; see also 
Fig.~\ref{edge_structure_fig}(a)).
The step height estimated from the line profile (not shown here) is 0.35 nm,
which is close to the graphite layer spacing ($= 0.335$ nm).
As seen in the figure, the edge is probably not a perfect straight line but 
has a slight irregularity.
In other words, it might consist of dominant zigzag edges and 
a small fraction of armchair edges.
Two types of superstructures are seen only on the upper terrace 
depending on the lateral position.
One is the $(\sqrt{3} \times \sqrt{3}) R 30^{\circ}$ superstructure and
the other is the honeycomb superstructure which consists of six B-site atoms.
Typical spatial extensions of the superstructures are 3$-$4 nm from the edge.
The superstructure pattern did not depend on the bias voltage in a range
between $+0.05$ and $+1.0$ V.
This is probably the first STM image showing clearly the 
superstructures near the zigzag edge.

Figure~\ref{armchair_superstructure_fig} is an STM image obtained near an 
armchair step edge.
The edge type is identified from an angle ($=90^{\circ}$) between the edge 
(the dashed line)
and an atomic row of B-site atoms (the long dashed line).
As was already reported by other workers~\cite{armchair_stm}, both the
$(\sqrt{3} \times \sqrt{3}) R 30^{\circ}$ and the honeycomb superstructures
are observed extending over 3$-$4 nm from the armchair edge.
Therefore, it can be concluded that these superstructures appear 
both on the terraces with zigzag and armchair edges.

\begin{figure}
\begin{center}
\includegraphics[width=4.5cm]{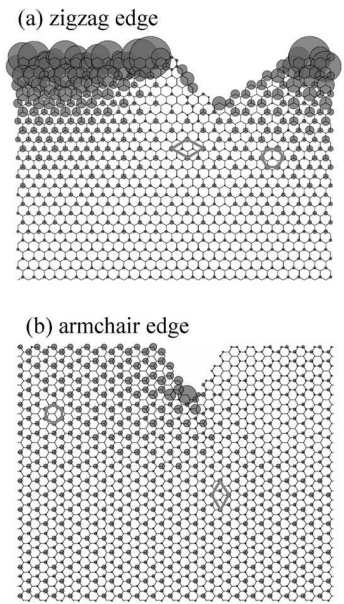}
\caption{Two examples of edge pattern used in the 
LDOS calculations:
(a) a zigzag edge with a small amount of armchair edges and (b) an armchair 
edge with a small amount of zigzag edges.
The radii of the circles plotted on the B-sites represent integrals of the 
calculated LDOS
in an energy range between $E_F$ and $+100$ meV.
The diamond and hexagon represent the same meanings as in 
Fig.~\ref{zigzag_superstructure_fig}.}
\label{mixture_edge_cal_fig}
\end{center}
\end{figure}

\begin{figure}
\begin{center}
\includegraphics[width=5.5cm]{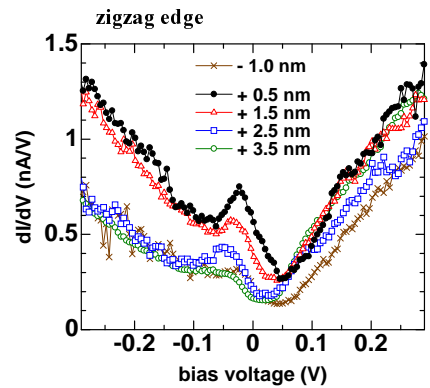}
\caption{$dI/dV$ vs. $V$ curves measured near a zigzag 
edge ($T=77$ K, in UHV).
The numbers denoted are distances from the edge ($d = 0$).}
\label{zigzag_sts_fig}
\end{center}
\end{figure}

\begin{figure}
\begin{center}
\includegraphics[width=5.5cm]{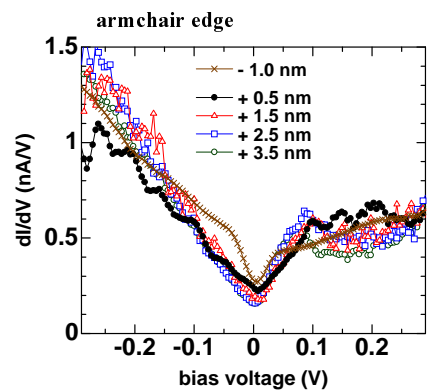}
\caption{$dI/dV$ vs. $V$ curves measured near an 
armchair edge ($T=77$ K, in UHV).
The numbers denoted are distances from the edge ($d = 0$).}
\label{armchair_sts_fig}
\end{center}
\end{figure}

We now compare the experimental STM images with those simulated by 
calculations based on 
the density-functional derived non-orthogonal tight-binding model
\cite{Frauenheim98}.
In the case of the perfect zigzag edge, the calculations indicate that the 
electronic states are
localized in the vicinity of the edge (decay length $\approx 0.5$ nm) and
no superstructures appear anywhere.
On the other hand, in the case of the perfect armchair edge, the calculations 
do not predict any
localized states at the edge but instead a honeycomb superstructure persisting
far beyond 5 nm from the edge.
These calculated results for the perfect edges are inconsistent 
with our STM observations.

Hence, we have calculated the LDOS near the zigzag (armchair) edges which are 
mingled
with small amounts of the armchair (zigzag) edges.
Two examples of such edge patterns are shown in 
Fig.~\ref{mixture_edge_cal_fig}(a)(b),
where integrals ($I_{cal}$) of the calculated LDOS in an energy range between 
$E_F$ and $+100$ meV
are represented by radii of the circles plotted on the B-sites.
Note that $I_{cal}$ should be proportional to the local tunnel currents at 
$V = +100$ mV.
As is seen in Fig.~\ref{mixture_edge_cal_fig}(a), both the $(\sqrt{3} 
\times \sqrt{3}) R 30^{\circ}$
and honeycomb superstructures appear on the terrace with the zigzag edges 
slightly mingled with the armchair edges.
The superstructures have decay lengths of about 4$-$5 nm from the edge
and rather complicated distributions in the parallel direction to the edge, 
which is consistent with the observation.
Although the atomic arrangement of the zigzag edge in the experiment
is not resolved clearly, the appearance of the two superstructures is 
strongly indicative
that the edge in Fig.~\ref{zigzag_superstructure_fig} is admixed with a 
small amount of the armchair edges.

The same is true for the other way around. Admixing of a small amount 
of the zigzag edges
to an armchair edge can explain the experimental coexistence of the two 
superstructures.
However, such admixing does not seem to explain the short decay lengths of 
the observed superstructures.
The predicted long persistence of the honeycomb 
superstructure near the armchair edges is absent in
Fig.~\ref{armchair_superstructure_fig}. This is probably 
due to the three dimensional 
character of the experimental system, 
which is not fully taken into account in the present calculations.
In any case, it is highly desirable 
to identify detailed edge shapes with atomic 
resolutions in the future.

Next, we show STS data taken near monoatomic step edges on ZYX 
obtained at 77 K in UHV.
We chose the scan directions parallel to the edges, and $dI/dV$ curves
taken at fixed distances ($d$) from the edges were averaged 
to increase the signal to 
noise ratio.
Observed corrugation amplitudes in the vicinity of the edges were rather 
position dependent both for the zigzag and armchair edges. This could be an 
additional indication of the mingling of the two types of edges. 
We took the spectroscopy data in Fig.~\ref{zigzag_sts_fig} 
(\ref{armchair_sts_fig}) at a position 
where the corrugation amplitude was the largest 
(smallest), based on the calculated results shown 
in Fig.~\ref{mixture_edge_cal_fig}.
Fig.~\ref{zigzag_sts_fig} shows tunnel spectra at different distances from 
a zigzag edge.
A small but clear peak appears around $V = - 20$ mV
and grows as the tip approaches the edge on the terrace ($d > 0$).
Note that the peak suddenly disappears when we move across the edge ($d = 
-1.0$ nm).
A decay length of this structure, estimated by subtracting a smoothed base 
line from the spectra, is
$1.6 \pm 0.4$ nm which is in agreement with that (1.2 nm) for $I_{cal}$ 
in Fig.~\ref{mixture_edge_cal_fig}(a).
Since the tunnel current was unstable at $|d| < 0.5$ nm for some reason, we 
could not obtain
reliable spectra right on the edge.
It should also be noted that the definition of $d = 0$ is somewhat 
arbitrary ($\pm 0.5$ nm)
because of the absence of the atomic resolution in that region.

In contrast to the zigzag edge, we obtained qualitatively different spectra 
near an armchair edge
as is shown in Fig.~\ref{armchair_sts_fig}, where such a LDOS peak 
is not observed within the experimental scatters.
Thus we conclude that the LDOS peak observed just below $E_F$ in 
Fig.~\ref{zigzag_sts_fig}
corresponds to the edge state that has been theoretically predicted to exist
only for the zigzag edge \cite{fujita1,miyamoto}.
A recent first-principles calculation for a graphite ribbon with zigzag edges of 3 nm wide
on an infinite graphene sheet \cite{tagami} indicates that the LDOS peak 
due to the edge state
is located below $E_F$ by a few tens mV.
This is in reasonable agreement with the peak in the tunnel spectra 
obtained in the present experiment.

\begin{acknowledgments}
One of us (H.F.) thanks the late Mitsutaka Fujita for stimulating his 
interest in
the graphite edge state.
The authors are grateful to H. Akisato for useful comments on this manuscript.
This work was financially supported by Grant-in-Aid for Scientific Research 
from
from MEXT, Japan and ERATO Project of JST.
One of us (T.M.) acknowledges JSPS Research Fellowship for Young Scientists.
\end{acknowledgments}

\newpage 

\end{document}